\documentstyle[multicol,aps,epsf]{revtex}
\begin{document}
\draft
\title{Domain Walls and Phase Transitions in the Frustrated 
Two-Dimensional XY Model}
\author{Colin Denniston$^{1,2}$ and Chao Tang$^2$}
\address{$^{1}$Department of Physics, Princeton University, 
Princeton, New Jersey 08544}
\address{$^{2}$NEC Research Institute, 4 Independence Way, Princeton,
New Jersey 08540}

\date{\today}
\maketitle

\begin{abstract}
We compare the critical properties of the two-dimensional (2D) $XY$ model in a 
transverse magnetic field with filling factors $f=1/3$ and $2/5$. 
To obtain a comparison with recent experiments, we
 investigate the effect of weak quenched bond disorder for $f=2/5$.  A 
finite-size scaling analysis of extensive Monte Carlo simulations strongly
suggests that the critical exponents of the phase transition for $f=1/3$ and 
for $f=2/5$ with disorder, fall into the 2D Ising model universality class.
Studying the possible domain walls in the system provides some 
explanations for our results.
\end{abstract}

\begin{multicols}{2}
The frustrated $XY$ model provides a convenient framework to study a
variety of fascinating phenomena displayed by numerous physical systems. 
One experimental realization of this model is in two-dimensional
arrays of Josephson junctions and superconducting wire networks
\cite{jja,us,Sean}.  A perpendicular magnetic field induces a finite
density of circulating supercurrents, or vortices, within the array.  The
interplay of two length scales -- the mean separation of vortices and the
period of the underlying physical array -- gives rise to a wide variety
of interesting physical phenomena.  Many of these effects show up as
variations in the properties of the finite-temperature superconducting
phase transitions at different fields.  Recent and ongoing experiments have
measured the critical exponents in superconducting arrays \cite{Sean}, opening 
the opportunity to do careful comparison of theory and experiment.  In this 
Letter we examine the critical properties of the 2D $XY$ model for two 
different values of the magnetic field in the densely frustrated regime 
($f \gg 0$) and in the presence of disorder.

The Hamiltonian of the frustrated $XY$ model is
\begin{equation}
{\cal H} = - \sum_{\langle ij \rangle} J_{ij} 
\cos(\theta_i-\theta_j-A_{ij}),
\label{ham}
\end{equation}
where $\theta_j$ is the phase on site $j$ of a square $L \times L$ lattice and
$A_{ij}=(2\pi/\phi_0)\int^j_i {\bf A} \cdot d{\bf l}$ is the integral of the
vector potential from site $i$ to site $j$ with $\phi_0$ being the flux 
quantum. The directed sum of the $A_{ij}$ around an elementary plaquette 
$\sum A_{ij}=2\pi f$ where $f$, measured in the units of $\phi_0$, is the
magnetic flux penetrating each plaquette due to the uniformly applied
field.  We focus here on the cases $f=p/q$ with $p/q=1/3$ and $2/5$.    

\begin{figure}
\narrowtext
\centerline{\epsfxsize=3.2in
\epsffile{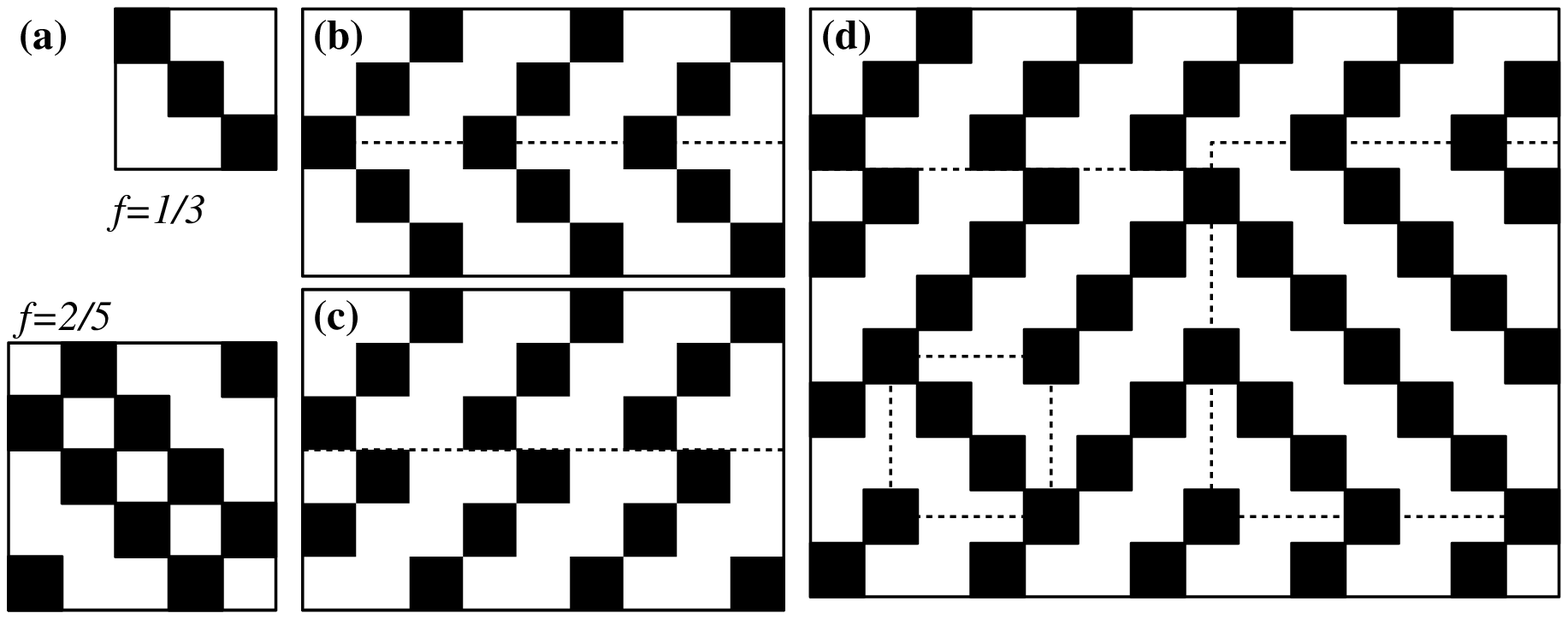}}
\vskip 0.1true cm
\caption{Fluxoid pattern for (a) unit cells of $f={1 \over 3}$ and 
$f={2 \over 5}$, and domain walls for $f={1 \over 3}$ (b) herringbone wall, 
(c) shift-by-one wall, and (d) shift-by-one wall 
branching into two herringbone walls (a vortex is shown as a dark square).}
\label{flux_walls}
\end{figure}

A unit cell of the ground state fluxoid pattern for these $f$ is 
shown in Figure~\ref{flux_walls}(a)\cite{TeitJay}.  The pattern consists of 
diagonal stripes composed of a single line of vortices for
$f={1\over 3}$ and a double line of vortices for $f={2 \over 5}$.
(A vortex is a pla\-quette with unit fluxoid occupation, 
ie. the phase gains $2 \pi$ when going around the pla\-quette.)  
The stripes shown in Figure~\ref{flux_walls}(a) can sit on $q$ sub-lattices, 
which we associate with members of the $Z_q$ group.
They can also go along either diagonal, and we associate these two 
options with members of the $Z_2$ group.
A common speculation for commensurate-incommensurate 
transitions and the frustrated $XY$ model is that the transition should
 be in the universality class of the q-state (or 2q-state) Pott's model.  
We find that this is not the case because domain walls 
between the different states vary considerably in both energetic and 
entropic factors.  

Table~\ref{wall_energies} lists the energy per unit length $\sigma$ for 
straight domain walls between the various ground states at zero temperature.  
We also numerically calculated the energy of domain 
walls that are not straight.  Closed domains, like those seen in the
simulations, of linear dimension $L$ from 10 to 60 unit cells in a system of 
size 120x120, have energies that scale linearly in $L$ to very high 
accuracy.  Examining two domains as a function of their distance apart shows 
only short range forces between them which fit an exponentially decaying 
function of distance very well\cite{us2}.  This strongly indicates that we can 
treat the energy of these domains as being linear in $L$. The other type of 
excitation (in addition to spin waves) is vacancy-interstitial pairs.  Such 
pairs have logarithmic interactions and can undergo a Kosterlitz-Thouless (KT) 
transition \cite{KTrefs}.  First, we focus on domain wall excitations.

The fluxoid pattern for the two lowest energy walls at $f$ $=$ ${1\over 3}$ is 
shown in Figure~\ref{flux_walls}.  One can see from the figure that 
a shift wall can be viewed as two adjacent, or {\it bound} herringbone walls. 
For $f={1 \over 3}$ the energy of two herring\-bone walls is less than that of 
a single shift wall and hence, the shift walls are unstable and break up into 
herringbone walls. As a result, we confine our discussion of the 
$f={1\over 3}$ case to the herringbone walls as other walls should not be 
present at large length scales.  The energy cost for dividing an $L \times L$ 
lattice into two domains separated by a solid-on-solid (SOS) wall stretching 
from one side of the system to the other is
\begin{equation}
{\cal H}_{single} \{z\}= b \sigma L+b \sigma \sum_k |z_k-z_{k-1}|.
\label{singleSOS}
\end{equation}
The height variables $z_k$, take on integer values ($b=3$ is the
shortest length segment).  The partition
 function, ${\cal Z}=\sum_{\{z_k\}}\exp (-{\cal H}/T)$ can be evaluated 
either by the transfer matrix method or recursively.  The interfacial free 
energy per column \cite{Forgacsetal} is
$
{\cal F}=T \ln [ e^{b \sigma/T}\tanh(b \sigma/(2T))].
$
The zero crossing of ${\cal F}$ gives an estimate of the critical 
temperature. Plugging in the values for the $f={1 \over 3}$ herringbone wall 
gives $T_c=0.19 J$, in remarkable agreement with the value $T_c=0.22 J$ 
found in the Monte Carlo simulations described below.  
Being similar to Ising walls, herringbone walls cannot branch into other 
herringbone walls, thus the set of possible domain wall configurations is 
similar to those in an Ising model.  We label the fraction of the system in
state $(s,j)$ as $m_{s,j}$, where $s=\pm1$ denotes the member of $Z_2$, and 
$j=1,2,3$ denotes the member of $Z_3$.  Below the transition, one state 
$(s,i)$ spans the system.  On this state sit fluctuating domains, bounded by 
herringbone walls, of each of the states $(-s,1), (-s,2),$ and $(-s,3)$ in 
equal numbers; so the $Z_3$ symmetry is broken for the $(s,j)$ states, but not 
for the $(-s,j)$ states.  As the transition is approached from below, the 
domains occupied by the $(-s,j)$ states grow, with smaller domains 
 of the $(s,j)$ states within them.  At the transition,
the $Z_2$ symmetry between the $\pm s$ states is restored and, as a result,
the $Z_3$ symmetry for the $(s,j)$ states is also restored.  

The Monte Carlo simulations used a heat bath algorithm with system sizes of 
$20 \le L \le 96$.  We computed between $10^7$ and $3\times 10^7$ 
Monte Carlo steps (complete lattice updates) with the largest fraction close to
 the transition temperature.  Data from different temperatures was 
combined and analyzed using histogram techniques \cite{FerrenburgSwend}.

If the largest fraction of the system is in state $(s,i)$, then we
have three Ising order parameters, 
$M_j=(m_{s,i}-m_{-s,j})/(m_{s,i}+m_{-s,j}), \, j=1\cdots 3.$ 
On average, these $M_j$ are the same so we just take the average as $M$.  To 
calculate the $m_{\sigma,i}$, we examine the 
Fourier transform of the vortex density $\rho_{k\pm}$ at the 
reciprocal lattice vectors ${\bf k_\pm}={\pi \over 3}(1,\pm1)$ of the ground 
state vortex lattices.  Starting from the definition of the Fourier transform, 
and using the vortex states given above, one finds 
$\rho_{k\pm}/\rho_g$ $=$ $m_{\pm 1,1}+m_{\pm 1,2}e^{i 2 \pi/3}+$ 
$m_{\pm 1,3}e^{-i 2 \pi/3}$, where $\rho_g$ is the modulus in the ground state.
In practice, $\rho_{k\pm}$ is reduced by small short-lived regions 
which don't quite match any of the six states.  Since this effect is the
same for all states, it cancels when calculating $M$. 
Using the real and imaginary parts of $\rho_{k\pm}$ in addition to 
$\sum_j m_{\pm1,j}$, calculated from the direct vortex lattice as in 
\cite{LeeLee}, we can find the five independent $m_{\sigma,j}$.

\begin{figure}
\centerline{\epsfxsize=3.2in
\epsffile{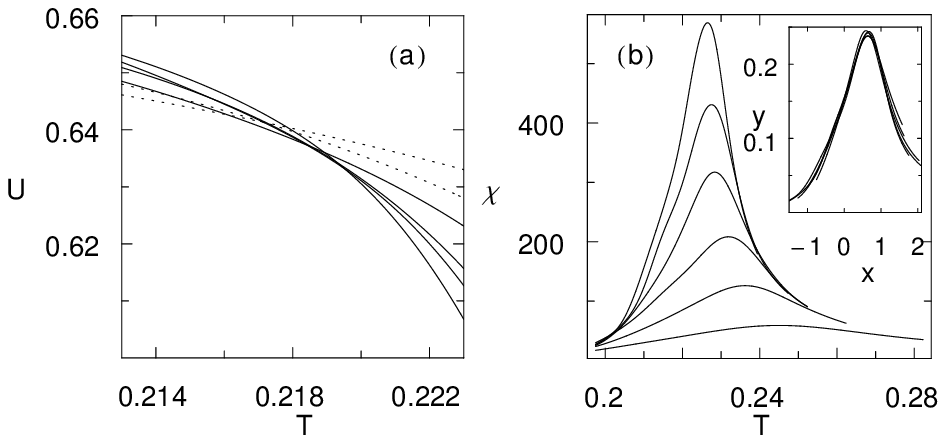}}
\vskip 0.1true cm
\caption{$f=1/3$ (a) Binder's cumulant $U$ vs $T$ for $L=36$ to $L=84$ (smaller $L$ shown as dotted lines), and (b) $\chi$ vs $T$ for $L=36$ to $L=84$ and
scaling collapse of this data (inset) where $x=(T-T_c)L^{1/\nu}$, 
$y=\chi L^{-\gamma/\nu}$, $\nu=1$, and $\gamma={7 \over 4}$.}
\label{Vchi}
\end{figure}

The transition temperature is located using Binder's cumulant \cite{Binder}, 
$U=1-\langle M^4 \rangle/(3 \langle M^2 \rangle^2)$, shown
in Figure~\ref{Vchi}(a).  For system sizes large enough to
 obey finite-size scaling, this quantity is size independent at the critical 
point.  From Fig.~\ref{Vchi} we find $T_c=0.2185(6)J$.  $T_c$ can also be 
determined from the scaling equation for the temperature at the peak of 
thermodynamic derivatives such as the susceptibility, 
$T_c(L)=T_c+a L^{-1/\nu}$. We find these other methods give $T_c$ in agreement
 with that from $U$.

\begin{figure}
\centerline{\epsfxsize=3.2in
\epsffile{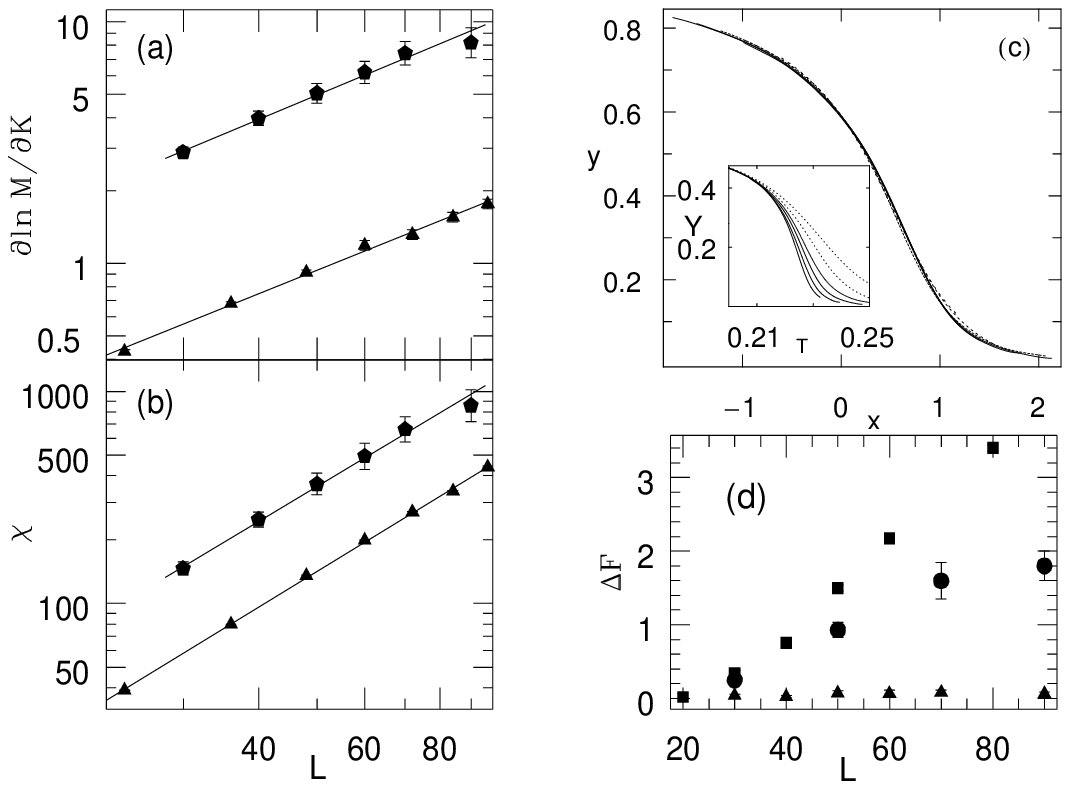}}
\vskip 0.1true cm
\caption{Finite size scaling plots for $f={1 \over 3}$ (triangles) and 
$f={2 \over 5}, \delta=0.1$ (pentagons): (a) logarithmic derivative of $M$ vs
 $L$, (b) $\chi$ vs $L$. (c) Scaling collapse of $Y$ (raw data in inset) where
$x=(T-T_c)L^{1/\nu}$, $y=Y L^{\beta/\nu}$, $\nu=1$, and $\beta={1 \over 8}$.
(d) Free energy barrier vs system size for $f={2 \over 5}$ and $\delta=0$ 
(squares), $\delta=0.05$ (circles) and $\delta=0.10$ (triangles).}
\label{scale}
\end{figure}

The exponents for the specific heat $\alpha$, the order parameter $\beta$, the 
susceptibility $\gamma$, and the correlation length $\nu$ describe the usual 
power-law singularities in the infinite system limit.
Finite size scaling \cite{FSS} at $T_c$ applied to 
$\partial \ln M / \partial K$ \cite{Chenetal} gives $1/\nu=1.007(25)$, and to
 the susceptibility $\chi$ gives $\gamma/\nu=1.743(20)$, and to $M$ gives 
$\beta/\nu=0.142(20)$ (these exponents are determined from the slopes of the 
lines shown in Fig.~\ref{scale}(a) and (b)).  This is in excellent agreement 
with the Ising values $\nu=1$, $\gamma={7 \over 4}$, and $\beta={1 \over 8}$.  
Fig.~\ref{Vchi} shows the collapse of the raw data onto the scaling function 
(inset) for $\chi$.
   
Two previous examinations of the $f={1 \over 3}$ case \cite{Faloetal} suggested
 a continuous transition but did not measure critical exponents.
Lee and Lee \cite{LeeLee} claim to find separate, closely spaced transitions, 
for the breaking of $Z_2$ and $Z_3$.  One explanation for their conflicting 
results comes from the small system sizes ($L \le 42$) used in their analysis. 
 Below the transition, if the dominant state is $(s,i)$, in small system sizes 
you often do not see all three of the $(-s,j)$ states in the system at the same
 time.  This can give the impression of separate 
transitions for small system sizes.  This impression is enhanced by the 
presence of a shoulder in the specific heat at intermediate system sizes 
\cite{LeeLee}.  For the larger system sizes, we see this shoulder merge with 
the main peak and for $L=84$ and $L=96$ it is no longer discernible.  
 
The helicity modulus $Y$ is the quantity most closely related to experimental 
measurements\cite{KTrefs}.  For $f \ne 0$, the scaling of the $I$-$V$ curves 
found in experiments is consistent with domain wall activation 
processes \cite{Sean}.
The theory of Nelson and Kosterlitz for the $f=0$ case predicts that $Y$ should
 come down in a characteristic square-root cusp and
then jump with the universal value, $2k_BT_{KT}/\pi$.  However, we find an
exceptionally good fit (Fig~\ref{scale}(c)) of our data to 
$Y-Y_0$ $=$ $L^{-\beta/\nu}$ ${\cal M}((T-T_c)L^{1/\nu})$ with $\nu=1$, 
$\beta$ $=$ ${1 \over 8}$, and $Y_0$ $=$ $0$.  Clearly, $Y$ is strongly 
renormalized from its bare value and attempting to fit scaling relations for 
the $f=0$ case 
\cite{LeeLee} without taking this into account seems questionable.  We see two 
possible interpretations of our result.  The first is that $Y$ only 
receives contributions from the ordered part of the lattice. 
So comparisons with the $f=0$ case should examine $Y_m=Y/M$.  
$Y_m\approx 0.58$ at the transition implying a larger than 
universal jump.  Alternatively, one can say that although $Y$ is brought 
down by the presence of fluctuations in $M$, it should still jump when it
crosses the universal value, $2k_BT/\pi$.  Extrapolating 
the observed behavior of $Y$ gives 
$Y_{L\rightarrow \infty}$ $=$ $a|T-T_c|^\beta$. 
This crosses the value of the universal jump at $T_{KT}-T_c \approx 10^{-6}$.
Although we do not see evidence for a jump (the best fit has $Y_0=0$), a 
difference in transition temperatures of $10^{-6}$ would not 
lead to any observable effects for the system sizes studied here. 

While for $f={1\over 3}$, herringbone walls are the only stable walls, this is
 not true for $f={2\over 5}$.  For 
$f={2\over 5}$ it is energetically favorable for two herringbone walls to bind 
and form a shift-by-one or shift-by-three wall.  Binding does, however, 
have an entropic cost.  To see if these walls are bound we consider the 
following model for two SOS walls:
\begin{eqnarray}
{\cal H}_{double} \{\Delta,z\}= 
	\sum_k \{(2b\sigma+u_\parallel\delta_{z_k,0})
	+b\sigma |z_k-z_{k-1}|\nonumber\\
	 +(2b\sigma+u_\perp \delta_{z_k,0}) \Delta_k
	+V_r(\{\Delta,z\})\}.
\label{doubleSOS}
\end{eqnarray}
$z_k$ is the separation of the walls ($z_k \geq 0$),
$\Delta_k$ is the number of vertical steps the two walls take in the same
direction in the k'th column ($-\infty < \Delta_k <\infty$).  $u_\parallel$
and $u_\perp$ are the binding energies parallel and perpendicular
to the wall.   At this stage we take $V_r=0$.  Summing over 
$\Delta_k$ leaves the partition function in the form of a transfer 
matrix: ${\cal Z}$ $=$ $\sum_{\{z_k\}} \prod_k T_{z_k}^{z_{k-1}}$.
Restricting $z_k-z_{k-1}$ to $0$ or $\pm 1$, we can derive the eigenvalues and 
eigenvectors of the matrix $\hat {\bf T}$ explicitly.  A ground state 
eigenvector $\psi_\mu(z)$ $=$ $e^{-\mu z}$, where $1/\mu$ is the localization 
length, or typical distance separating the lines, characterizes the bound state
 of the two lines.  $\mu=0$ defines the unbinding transition at $T_b$.  
Repeating this process numerically for the unrestricted case 
($z_k-z_{k-1}\geq 0$) gives $T_b=0.398 J$ for the shift-by-one walls and
$T_b=0.442 J$ for the shift-by-three walls.  The free energy for these walls
crosses zero before they unbind. Hence, at the transition we expect a 
branching domain wall structure similar to the $q\ge 5$ Pott's models where a 
first order phase transition occurs.

In their Monte Carlo simulations, Li and Teitel \cite{LiTeitel} observed 
hysteresis of the internal energy when the temperature was cycled around the 
transition and used this as an argument for a first order 
transition at $f={2\over 5}$.  The most direct indication of a first order 
transition is the presence of a free energy barrier between the 
ordered and disordered states which diverges as the system size increases 
\cite{LeeKosterlitz}.  The free energy as a function of energy is obtained
 using ${\cal F}_L(E)=-\ln P_L(E)$ where $P_L(E)$ is 
the probability distribution for the energy generated by Monte Carlo 
simulation of a $L\times L$ system. 
Figure~\ref{scale}(c) shows the growth in this barrier as the 
system size increases from $L=20$ to $80$ giving clear evidence for the 
first order nature of the transition.  Also, according to finite size 
scaling, the maximum of $C$ and $\chi$ should scale with $L^d$ for
first order phase transitions \cite{FSS}.  We find this to be the case and 
also obtain $T_c=0.2127(2) J$ \cite{us2}.  

We now consider the effects of disorder on the $f={2\over 5}$ phase transition.
Taking the couplings in the Hamiltonian (\ref{ham}) as 
$J_{ij}=J(1+\epsilon_{ij})$, the $\epsilon_{ij}$ are chosen randomly
from a Gaussian distribution with a standard deviation $\delta$.
Due to variations of the phase differences across the bonds, a specific 
realization of random bonds may favor a certain sub-lattice for the ground 
state, creating an effective random field. To quantify the effect, we placed 
the fluxoid configuration of the ground states down on $10\,000$ 
separate realizations of the disorder and allowed the continuous degrees of 
freedom (the phases) to relax and minimize the energy.
We find that the energy shifts from the $\delta=0$ case, and these shifts fit
a Gaussian distribution with mean $-0.5 \delta^2 L^2$ and standard
deviation $\delta L$.  The difference in energy between states which 
were degenerate in the clean system is the measure of the 
random field.  This difference centers on zero and has
a standard deviation of $0.75 \delta L$ for two states related
by a shift and $0.57 \delta L$ for two states with vortex rows
along opposite diagonals.  The effect of random fields on discrete degrees of 
freedom in 2d is marginal \cite{Imry-Ma-Aizenman-Wehr-Hui-Berker}. For $d>2$ 
there is a critical randomness above which random fields cause the formation of
 domains in the ground state of size $\sim \xi_{RF}$.  Aizenman and Wehr have 
shown that this critical randomness is zero in 2d 
\cite{Imry-Ma-Aizenman-Wehr-Hui-Berker}.  Yet, their result does 
not preclude the possibility that 
$\xi_{RF}$ is so large as to be unobservable in a finite sized sample.  Indeed,
 experiments on superconducting arrays have found apparent phase transitions, 
including scaling behavior \cite{Sean} in sample sizes of order 
$1000 \times 1000$.  In our simulations with disorder at $\delta \leq 0.1$,
 all systems had a low temperature state with the order parameter approaching 
unity.  We will, therefore, ignore the effects of random fields for 
$\delta \leq 0.1$ assuming that $\xi_{RF}$ is larger than the sample size.

At any coexistence point of the clean system, random {\it bonds} result in 
different regions of the system experiencing average couplings slightly above 
or below the critical coupling.  As a result, at any given temperature the 
system will predominantly prefer 
either the ordered or disordered state wiping out the coexistence region and 
leaving only a continuous transition \cite{Imry-Ma-Aizenman-Wehr-Hui-Berker}.  
It has been conjectured \cite{Chenetal} 
that critical random Potts models are equivalent to Ising models.  Kardar et 
al. \cite{KardarII} suggested a possible mechanism for this effect.  Their 
position space renormalization group approximation suggests that the 
probability of loop formation in the fractal interface of the clean system 
vanishes marginally at a transition dominated by random bonds.  The 
interface may have some finite width due to a froth of bubbles of different
phases, but under renormalization a linear critical interface is obtained and, 
hence, an Ising transition appears.  

The fluxoid configurations from our simulations suggest that for large
enough disorder, ($\delta>\delta_f$) the interface is really linear, not just 
in the renormalized sense.  $\delta_f$ can be estimated by placing a random 
potential $V_r$ in Eq.~\ref{doubleSOS}.  Ignoring the terms involving 
$\Delta_k$, one obtains the model for wetting in the presence of disorder, 
solved by Kardar\cite{Kardar} in the continuum limit.  
He obtained a new length scale due to randomness,
$
1/\kappa=2T^3 / K \delta^2
$
where K is the renormalized stiffness 
\cite{Forgacsetal}.
The unbinding transition is lowered and is now defined by the condition 
$\mu-\kappa=0$.  As $T_b$ decreases, it
eventually hits the transition temperature for the first order phase transition
 observed in the clean system.  At this point any branched domain wall 
structure is unstable.  This is just the last step in a process in which the 
effective linear interface becomes narrower as disorder increases.  In the 
vicinity of this ``final'' unbinding, the Ising-type behavior of the system 
should be readily visible at any length scale.

We have done a Monte Carlo analysis with bond disorder values of 
$\delta=0.05$ and $0.1$.  To calculate the average value of a thermodynamic 
quantity, we first calculate it for a given realization of the disorder and 
then do a configurational average over 10 to 15 realizations for 
$\delta=0.1$ and seven realizations for $\delta=0.05$. 
Figure~\ref{scale}(c) shows the free energy barrier for $f={2\over 5}$ as a 
function of system size in the for $\delta=0.05$, and $0.1$.  
For $\delta=0.05$, the barrier first grows with system size and then levels 
off.  At $\delta=0.1$ the free energy barriers are essentially zero, 
indicating a continuous transition and that the system sizes are large enough 
to apply finite size scaling.  Here, we follow the finite-size scaling methods
used in \cite{Chenetal}.

Figure~\ref{scale} shows the peak values of $\partial \ln M /\partial K$ and 
$\chi$ as a function of $L$. The slopes of these 
plots give $1/\nu=1.05(12)$ and $\gamma/\nu=1.70(12)$.  A similar analysis of
$\partial M/\partial K$ gives $(1-\beta)/\nu=0.94(10)$ \cite{us2}.
Within errors, these exponents are what one
would expect from an Ising model. Experiments at $f={2\over 5}$ \cite{Sean} 
also found a continuous transition and measured the critical exponents
$\nu=0.9(5)$ and the dynamic critical exponent $z=2.0(5)$, consistent with
an Ising transition.

In conclusion, we find that the nature and universality class of the phase
transitions are quite sensitive to the proximity of the binding transition for 
the lowest energy domain walls.  For $f=1/3$ the lowest energy walls are never 
bound and the transition is Ising-like.  For $f=2/5$ domain walls can lower 
their free energy by binding to each other, resulting in
 a first order phase transition.  Disorder weakens this binding and 
changes the transition to be continuous and Ising-like.  Our results are 
consistent with the continuous phase 
transition and critical exponents observed experimentally for 
$f=2/5$ \cite{Sean}. 

We thank M. Aizenman, P. Chandra, J.M. Kosterlitz, X.S. Ling, and D. Huse and 
for useful discussions.

\begin{table}
\narrowtext
\begin{tabular}{ccc}
{\it domain wall type} & \multicolumn{2}{c}{\it energy per unit length}\\
	       & $f=1/3$        & $f=2/5$ \\ 
\hline
herringbone    &  0.056737424 J &  0.086117262 J \\
shift-by-one   &  0.114199976 J &  0.158899286 J \\
shift-by-two   &  0.166666666 J &  0.166122315 J \\
shift-by-three &                & 0.147648594 J  \\
shift-by-four  &                & 0.198688789 J  \\
\end{tabular} 
\caption{Domain wall energies.}
\label{wall_energies}
\end{table}

\end{multicols}

\end{document}